\documentclass[letter,rmp,twocolumn,amsmath,amssymb]{revtex4}
\usepackage{graphicx}
\usepackage{dcolumn}
\usepackage{bm}

\begin{document}

\newcommand{\be}{\begin{equation}}
\newcommand{\ee}{\end{equation}}
\newcommand{\bes}{\begin{equation}}
\newcommand{\ees}{\nonumber\end{equation}}
\newcommand{\bea}{\begin{eqnarray}}
\newcommand{\eea}{\end{eqnarray}}
\newcommand{\uv}[1]{\mathbf{\hat{#1}}}
\newcommand{\curl}[1]
{\mathbf{\nabla}\times\mathbf{#1}}
\newcommand{\dvg}[1]
{\mathbf{\nabla}\cdot\mathbf{#1}}
\newcommand{\pd}[2]
{{{\partial #1} \over {\partial #2}}}
\newcommand{\pdt}[2]
{{{\partial^{2} #1} \over {\partial #2 ^{2}}}}

\newcommand{\m}[1]
{\mathbf{#1}}

\preprint{Draft}

\title{Time-domain measurement of driven ferromagnetic resonance}

\author{Y. Guan\email{yg2111@columbia.edu} and W. E. Bailey}
\affiliation{Materials Science Program, Department of Applied
Physics, Columbia University, New York, New York 10027}
\author{E. Vescovo, C.-C. Kao, and D. A. Arena}
\affiliation{National Synchrotron Light Source, Brookhaven
National Laboratory, Upton, New York 11973}

\date{\today}

\begin{abstract}

We present a time-resolved measurement of magnetization dynamics
during ferromagnetic resonance (FMR) in a single layer of
Ni$_{81}$Fe$_{19}$. Small-angle ($<$1$^{\circ}$) precession of
elemental Ni, Fe moments could be measured directly and
quantitatively using time-resolved x-ray magnetic circular
dichroism (XMCD) in transmission. The high temporal and rotational
sensitivity of of this technique has allowed characterization of
the phase and amplitude of driven FMR motion at 2.3 GHz, verifying
basic expectations for a driven resonance.

\end{abstract}

\maketitle

Ferromagnetic resonance (FMR) is a venerable topic in the study of
magnetism. In the modern technological context, resonance and
relaxation underpin the switching response of spin electronic
devices at 1 GHz and above.

In this Letter, we demonstrate a time- and element-resolved
measurement of ferromagnetic resonance in a single layer of
Ni$_{81}$Fe$_{19}$. Conventional, low-angle ($\sim$
0.1-1.0$^{\circ}$) FMR motion, driven with a continuous wave (CW)
low-power microwave field at 2.3 GHz, has been measured in the
time-domain using time-resolved x-ray magnetic circular dichroism
(XMCD).

Two innovations have allowed the rotational and spatial resolution
necessary for the measurement. High magnetic contrast is provided
by transmission geometry XMCD\cite{chen-prl, guan-jap}, the soft
x-ray equivalent of Faraday rotation. Cone angles could be
measured down to 0.1$^\circ$. Improved temporal resolution is
provided using phase-locked CW microwaves as an excitation source,
suppressing the effects of timing jitter present in pulsed
experiments\cite{silva-jap, bailey-prb}. Motional and phase
resolutions are an order of magnitude better than we achieved in
previous work\cite{bailey-prb} using pulsed step fields in XMCD
reflectivity.

In general, magnetization dynamics are described by the
Landau-Lifshitz (LL) equation\cite{landau}, given in SI as
\begin{equation}
{d\mathbf{M}\over dt} = - \mu_{0}\mid \gamma \mid \left(\mathbf{M}
\times \mathbf{H}\right) - {\lambda\over M_{s}^{2}}
\left(\mathbf{M} \times \mathbf{M} \times \mathbf{H} \right),
\end{equation}
where $\lambda$ is the LL relaxation rate in sec$^{-1}$.

The LL equation can be linearized for small rotations of $\vec{M}$
about $\vec{H}$ as
\begin{equation}
\pdt{\phi(t)}{t}+\lambda\pd{\phi(t)}{t}+\omega_{0}^{2}\phi(t) = 0,
\end{equation}
where $\omega_{0}^2=\mu_{0}^{2}\gamma^{2}H_{eff}(H_{eff}+M_{s})$
\cite{kittel}.

Free oscillations, describing the motion of this damped harmonic
oscillator about an equilibrium position, are the starting point
for most magnetooptical studies of spin dynamics\cite{silva-jap,
bailey-prb, crowell-jap, meckenstock-apl, Tamaru-prb}. Rotational
displacements about an equilibrium are described by $\phi$;
$\omega_{0}$ is is the circular frequency of ferromagnetic
resonance (FMR) and $2/\lambda$ is its characteristic relaxation
time.

If instead the motion is forced by a transverse ac field
$H_{y}(t)=H_{y0}\exp{(i\omega t)}$, the response is given as
\begin{equation}
\pdt{\phi(t)}{t}+\lambda\pd{\phi(t)}{t}+\omega_{0}^{2}\phi(t) = A \exp{(i\omega t)},
\end{equation}
where $A \approx \mu_{0}^{2}\gamma^{2}M_{s}H_{y0}$. Solving Eq.
(3) using $\phi(t)=\phi_{0}\exp{(i\omega t)}=\mid\phi_{0}\mid
\exp{[i(\omega t + \delta)]}$, then
\begin{equation}
\phi_{0} = {A
\over(\omega_{0}^2-\omega^2)^2+\lambda^2\omega^2}[(\omega_{0}^2-\omega^2)-i\omega\lambda].
\end{equation}
Thus, the phase $\delta$ and the amplitude of driven FMR precession can be expressed as:
\begin{equation}
\tan\delta = {-\lambda\omega\over\omega_{0}^2-\omega^2},
\end{equation}
\begin{equation}
\mid\phi_{0}\mid = {A
\over\sqrt{(\omega_{0}^2-\omega^2)^2+\lambda^2\omega^2}}.
\end{equation}

These relationships can now be tested directly.  Moreover, using
{\it in-situ} FMR (microwave absorption) measurement, the damping
$\lambda$ can be estimated directly through the field linewidth,
allowing for a parameter-free comparison with Eqs. (5) and (6).

FMR absorption is given by the imaginary part of the
susceptibility, $\chi^{''}$, along the rf driving field, $H_{y}$,
according to Eq. (4). The field-swept resonance linewidth has
half-power points at $\mu_{0}\Delta H_{1/2} = {2
\alpha\omega/\gamma}$, directly proportional to the dimensionless
damping constant $\alpha$. Between these half-power points, the
phase lag $\delta$ of $\phi(t)$ with respect to the drive field
goes through a change of 90$^{\circ}$ according to Eq. (5).

In lock-in (derivative) detection of microwave absorption, the
inflection points of the Lorenzian line shape are more easily
seen. These have a width $\mu_{0}\Delta H_{pp} =
(2/\sqrt{3}){\alpha\omega/\gamma}$\cite{heinrich}. While $\alpha =
{\lambda/(\mu_{0}M_{s}\gamma)}$ for low damping ($\alpha\ll 1$),
$\lambda$ can thus be expressed as
\begin{equation}
\lambda = {{\sqrt{3}\over{2}}{\mu_{0}^2\gamma^2M_{s}\Delta
H_{pp}\over{\omega}}}.
\end{equation}
Then, the following relationship can be derived as
\begin{equation}
\sqrt{3}\Delta H_{pp} = \Delta H_{1/2} = {1\over{\sqrt{3}}}\Delta
H_{{1\over{2}}\mid\phi_{0}\mid},
\end{equation}
where $\Delta H_{{1\over{2}}\mid\phi_{0}\mid}$ denotes the
linewidth defined by the half-amplitude points.

Time-resolved XMCD (TR-XMCD) measurements of magnetization motion
during FMR precession were carried out at Beamline 4-ID-C of
Advanced Photon Source in Argonne, IL. The circular dichroism
signal was obtained in transmission, using photon helicity
$\vec{\sigma}$ switching ($\vec{\sigma}$, with an incident angle
of 38$^\circ$ from normal, and with in-plane projection along
$\uv{y}$) at the elliptical undulator for fixed static applied
field $H_{B}$. The transmitted intensity was read at a soft x-ray
sensitive photodiode and normalized to an incident intensity at a
reference grid.

Transmission measurements require an x-ray transparent sample and
substrate. For time-resolved measurements, x-ray transparent RF
field delivery system is also necessary. The
Ni$_{81}$Fe$_{19}$(25nm)/Cu(2nm) thin film sample was deposited
onto a Si/Si$_{3}$N$_{4}$ membrane using UHV magnetron sputtering
at a base pressure of 4$\times$10$^{-9}$ Torr. The sample was
placed in the center of a hollow microwave resonator. Uniform
precession of the magnetization was excited at 2.3 GHz by a CW
low-power microwave field, synchronized with variable delay to APS
x-ray photon bunches (88 MHz). Microwave absorption was measured
{\it in-situ}, using standard lock-in techniques, detecting
reflected power at the resonator. Orthogonal Helmholtz coils were
used to apply longitudinal bias field $H_{B}\uv{x}$ or transverse
bias field $H_{T}\uv{y}$.

Element-specific XMCD hysteresis loops were taken as a function of
transverse bias field $H_{T}$ to obtain a calibration for
magnetization angle $\phi$. Photon energies were set to the
$L_{3}$ peaks for Fe (707.5 eV) and Ni (852.0 eV) to measure Fe
and Ni XMCD signals, respectively. The saturation values of XMCD
signals are taken to be $\phi_{Fe}=\phi_{Ni}=\pm 90^\circ$.

$L_{2,3}$-edge XAS and MCD spectra have been measured in
transmission for both Fe and Ni in Ni$_{81}$Fe$_{19}$.
High-quality spectra are obtained here, as shown in our previous
work at NSLS, Beamline U4B\cite{guan-jap}. Fig. 1(a) shows Fe
transmission XAS spectra for both photon helicity directions, with
the difference. Corresponding spectra for Ni are shown in Fig.
1(b).

\begin{figure}[h]
\includegraphics[width=\columnwidth]{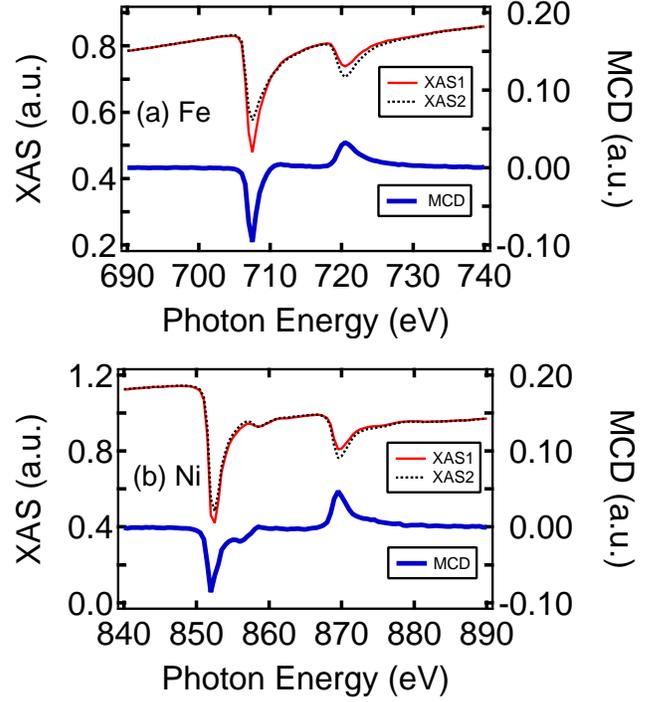}
\caption{(a)$L_{2,3}$-edge transmission XAS and MCD spectra of Fe
in Ni$_{81}$Fe$_{19}$;(b)$L_{2,3}$-edge transmission XAS and MCD
spectra of Ni in Ni$_{81}$Fe$_{19}$.}
\end{figure}

Time- and element-resolved XMCD measurements of magnetization
precession at resonance are presented in Fig. 2. XMCD signals were
taken as a function of delay time and converted into
time-dependent elemental magnetization angles $\phi_{Fe}(t)$ and
$\phi_{Ni}(t)$ for Fe and Ni, respectively\cite{bailey-prb}.
Precessional oscillations are clearly seen. Fe and Ni moments are
found to precess together within instrumental resolution, improved
here to $\pm$2 ps and $<$0.1$^\circ$.

\begin{figure}[h]
\includegraphics[width=\columnwidth]{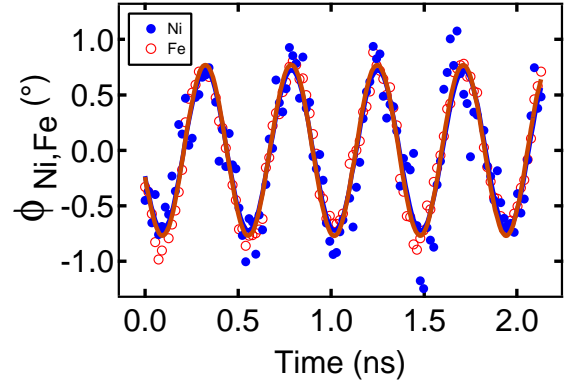}
\caption{TR-XMCD measurement of Fe and Ni magnetization precession
at resonance, 37 Oe at 2.3 GHz. Solid lines are sinusoidal fits of
the Fe and Ni data sets separately.}
\end{figure}

Time-resolved XMCD measurements of magnetization precession off
resonance are presented in Fig. 3. Applied fields were selected
according to {\it in-situ} measured FMR spectra (Fig. 4(a)),
spanning the resonance condition H$_{res}$ (37 Oe) to $\sim
4\times\Delta H_{pp}$ off resonance (5 Oe). A clear variation in
the amplitude of driven FMR motion $\phi_{Fe}(t)$, and its phase,
compared with the RF excitation field, can be seen.

\begin{figure}[h]
\includegraphics[width=\columnwidth]{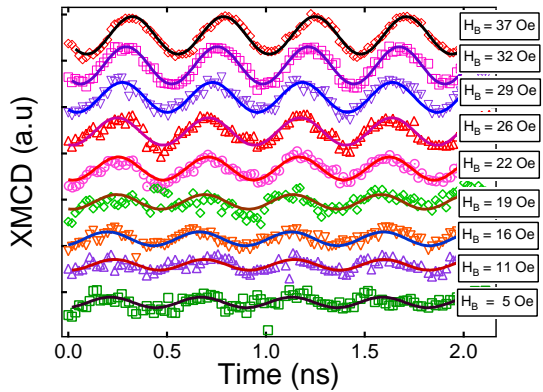}
\caption{TR-XMCD measurement of Ni$_{81}$Fe$_{19}$ magnetization
precession off resonance at Fe $L_{3}$ edge. Solid lines are
sinusoidal fits.}
\end{figure}

The key result of this letter is presented in Fig. 4(b). Verifying
basic expectations of a driven resonance, we can clearly see a
90$^{\circ}$ phase shift generated through the adjustment of
$\omega_{0}$ (through longitudinal bias field $H_{B}$) to $\ll
\omega$, and a Lorentzian variation of the precessional amplitude.
Both behaviors are in excellent agreement with the linearized
model (Eqs. (5) and (6)). No empirical parameters have been used
apart from $H_{y0}$; $\lambda$ is estimated as 1.80 GHz directly
from the {\it in-situ} measured FMR spectra (Fig. 4(a)) using Eq.
(7). We verify directly the relationship in Eq. (8), with
$\sqrt{3}$ separating the peak-to-peak, 1/2 power (90$^\circ$
phase shift), and 1/2 amplitude linewidths.

\begin{figure}[h]
\includegraphics[width=\columnwidth]{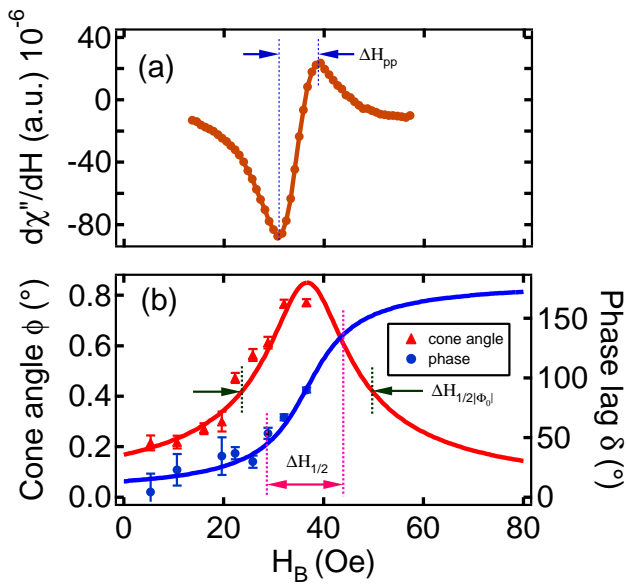}
\caption{(a){\it In-situ} measurement of FMR spectra of
Ni$_{81}$Fe$_{19}$ by microwave absorption at 2.3 GHz;
(b)Measurement of phase and amplitude of driven FMR precession in
Ni$_{81}$Fe$_{19}$ by TR-XMCD. Solid lines are the corresponding
theoretical simulations from Eqs. (5) and (6).}
\end{figure}

In conclusion, we have measured driven ferromagnetic resonance
(FMR) precession in the time domain using time-resolved XMCD.
Precessional phase and amplitude, lumped together in microwave
absorption measurement, could be measured directly, providing a
vivid illustration of damped oscillator behavior.

The authors thank Carl Patton for a critical reading of this
manuscript, and David J. Keavney (APS) for beamline support. This
work was partially supported by the Army Research Office with
Grant No. ARO-43986-MS-YIP, and the National Science Foundation
with Grant No. NSF-DMR-02-39724. Use of the Advanced Photon Source
was supported by the U.S. Department of Energy, Office of Science,
Office of Basic Energy Sciences, under Contract No.
W-31-109-Eng-38.

\end{document}